# LEARNING-BASED MODELS FOR BUILDING USER PROFILES FOR PERSONALIZED INFORMATION ACCESS

| | | |
|---|---|---|
| Minyar Sassi Hidri | Computer Department, Deanship of Preparatory Year and Supporting Studies, Imam Abdulrahman Bin Faisal University, Dammam, Saudi Arabia | mmsassi@iau.edu.sa |

## ABSTRACT

| | |
|---|---|
| Aim/Purpose | This study aims to evaluate the success of deep learning in building user profiles for personalized information access. |
| Background | To better express document content and information during the matching phase of the information retrieval (IR) process, deep learning architectures could potentially offer a feasible and optimal alternative to user profile building for personalized information access. |
| Methodology | This study uses deep learning-based models to deduce the domain of the document deemed implicitly relevant by a user that corresponds to their center of interest, and then used predicted domain by the best given architecture with user's characteristics to predict other centers of interest. |
| Contribution | This study contributes to the literature by considering the difference in vocabulary used to express document content and information needs. Users are integrated into all research phases in order to provide them with relevant information adapted to their context and their preferences meeting their precise needs. To better express document content and information during this phase, deep learning models are employed to learn complex representations of documents and queries. These models can capture hierarchical, sequential, or attention-based patterns in textual data. |
| Findings | The results show that deep learning models were highly effective for building user profiles for personalized information access since they leveraged the power of neural networks in analyzing and understanding complex patterns in user behavior, preferences, and user interactions. |

| Recommendations for Practitioners | Building effective user profiles for personalized information access is an ongoing process that requires a combination of technology, user engagement, and a commitment to privacy and security. |
| --- | --- |
| Recommendations for Researchers | Researchers involved in building user profiles for personalized information access play a crucial role in advancing the field and developing more innovative deep-based networks solutions by exploring novel data sources, such as biometric data, sentiment analysis, or physiological signals, to enhance user profiles. They can investigate the integration of multimodal data for a more comprehensive understanding of user preferences. |
| Impact on Society | The proposed models can provide companies with an alternative and sophisticated recommendation system to foster progress in building user profiles by analyzing complex user behavior, preferences, and interactions, leading to more effective and dynamic content suggestions. |
| Future Research | The development of user profile evolution models and their integration into a personalized information search system may be confronted with other problems such as the interpretability and transparency of the learning-based models. Developing interpretable machine learning techniques and visualization tools to explain how user profiles are constructed and used for personalized information access seems necessary to us as a future extension of our work. |
| Keywords | personalized IR, user profile, deep learning, recurrent neural networks, artificial neural networks, convolutional neural networks |

# Introduction

Information retrieval (IR) is a field of study and practice that involves the process of obtaining information from a large repository or dataset. The primary goal of information retrieval is to retrieve relevant information in response to a user's query. This field is integral to various applications, including search engines, databases, digital libraries, and more (Bassani, 2022; Büttcher et al., 2010; Chen & Kuo, 2000; Croft et al., 2009; Ibrihich et al., 2022; Lal et al., 2016; Roberts et al., 2017).

In the context of IR, the matching phase is pivotal for determining the relevance of documents to a user query. This phase involves comparing the query (user's information need) with the documents in the information system to identify and rank relevant documents. An effective matching phase helps in reducing the time and computational resources required for retrieval. By efficiently filtering out irrelevant documents early in the process, the system can focus on analyzing and ranking a smaller subset of potentially relevant documents.

Techniques such as the vector space model (SVM), keyword matching, statistical methods, or more advanced natural language processing (NLP) are commonly employed to express document content and information during this phase, enabling effective retrieval of relevance of relevant documents (Alsaif, Sassi Hidri, Eleraky, et al., 2022; Alsaif, Sassi Hidri, Ferjani, et al., 2022; Stathopoulos et al., 2023). They are used to assess the relevance of retrieved information to the user's query (Apostolico & Galil, 1997; Chen & Kuo, 2000; Garcés et al., 2006; Qi et al., 2020; Stathopoulos et al., 2023; Thakare & Dhote, 2013; Vijayarani & Janani, 2016; Zhu et al., 2023).

However, a significant challenge arises due to vocabulary differences between user queries and document content. Users may use different terms or phrases than those present in the documents they seek, leading to mismatches and potentially relevant documents being overlooked. This discrepancy can hamper the effectiveness of traditional matching techniques and hinder the retrieval of truly relevant information. As a result, selecting the information that best meets the user's needs rather than searching for information is the user's problem.

To address this issue, an innovative approach is proposed to bridge the gap between user queries and document vocabulary. One promising direction involves leveraging optimal deep learning architectures (Ferjani et al., 2022; Khoei et al., 2023) that offer a promising avenue for mitigating the impact of vocabulary differences in IR while enabling more accurate matching between user queries and document content. By learning representations that transcend individual terms, the proposed models effectively handle vocabulary mismatches and facilitate personalized information access.

The remainder of the article is organized as follows: The second section highlights related work on personalized information access and user profile modeling. The third section presents our motivation and methodology employed. The fourth section presents how data is data preprocessed and represented. Computational results are presented in the fifth section. The sixth section provides conclusions and highlights future directions.

## PERSONALIZED INFORMATION ACCESS AND USER PROFILE MODELING: RELATED WORK

Adapting, personalizing a document, or an application for a particular user requires a more elaborate description of the user and his representation as a full-fledged object of the system. This representation of the user aims to provide the system with the means to make the desired adaptations to evaluate the relevance of available objects (documents, web pages, etc.) or to help the system make choices (Gauch et al., 2007; Sowbhagya et al., 2022). Personalization of the information access process consists of integrating or exploiting the user profile in the information access chain (Lin et al., 2022). Its fundamental goal is to restore, at the top of the list of results, documents that interest the user in their search, in other words, which seem most similar to their profile.

Recent studies in personalized information access and user profile modeling have explored various innovative approaches and techniques to enhance the effectiveness and relevance of personalized recommendations (Farid et al., 2018; Lei et al., 2023; Oguntola & Simske, 2023; Purificato et al., 2024; Yan et al., 2023; Yang & Fang, 2013).

In contextual personalization, the focus is on incorporating contextual information, such as location, time, device, and user activity, into user profiles to improve the relevance of recommendations. Context-aware models dynamically adapt recommendations based on the user's current context, leading to more personalized and timely suggestions (Oguntola & Simske, 2023; Purificato et al., 2024; Yang & Fang, 2013).

With the increasing availability of multimedia content, recent studies have investigated techniques for personalized recommendation systems that can handle diverse data types, including text, images, audio, and video (Farid et al., 2018; Lei et al., 2023; Yan et al., 2023).

Multi-modal approaches enable a richer understanding of user preferences and interests, leading to more accurate and engaging recommendations.

To address concerns about algorithmic fairness and bias in personalized recommendation systems, recent research has focused on developing fairness-aware models and evaluation metrics (Gao et al., 2022; Lalor et al., 2024; Zhang et al., 2023). These studies aim to mitigate bias and ensure equitable treatment across diverse user groups, promoting diversity and inclusivity in personalized recommendations.

By leveraging these recent studies and advancements in personalized information access and user profile modeling, researchers developed effective approaches for integrating the profile into at least one of the phases of the information access process such as enrichment of queries, filtering of results, or reclassification of results (Belkin & Croft, 1992; Gauch et al., 2007; Koutrika & Ioannidis, 2004; Shu et al., 2020).

In the query enrichment approach, the service consists of enriching the user's query with a set of predicates contained in their profile. Query enrichment therefore uses the user's profile to reformulate their query by integrating elements defined in their profile (preferences, center of interest, etc.). The most successful method is that of Koutrika and Ioannidis (2004). User profiles are based on weighted predicates in this method. A predicate's weight expresses how important it is to the user. Real values between 0 and 1 are used.

In the filtering results approach, a user profile must specify the user's preferences in the filtering task. After that, incoming papers are compared to this profile to see which ones could be of interest to that specific user (Belkin & Croft, 1992). The system selects documents based on the user profile that it thinks the user would find interesting when fresh documents are received. The user specifies both relevant and non-relevant documents using a "Relevance feedback" procedure. With this data, the system modifies the user profile description to match the updated preferences. The advantage of filtering results is its simplicity because it does not require any modification to the operation of the information providers. All processing is done after the query is executed. The disadvantages are the volume of data exchanged between the server and the client and the risk of eliminating relevant elements.

In a personalized IR system that uses this approach, the system sends a query to a search engine, receives results, and then re-sorts the results based on their similarity to the user profile (Shu et al., 2020). In result reclassification, the user profile was utilized by the authors in (Gauch et al., 2007) to reorder the *ProFusion* meta-search engine's results. The original engine result, the degree of similarity between the result and the related concepts, and the user's interest in these ideas (represented by the weights of these concepts) are used to compute a new score, $new_{wt_r}$ for each result $r$ (returned document) of this meta-engine:

$$new_{wt_r} = wt_r \left( 0.5 + \frac{1}{4}\sum_{l=1}^{4} u_{crl} \right) \tag{1}$$

where:

- $wt_r$ is the original score calculated by the search engine for the result r.
- $u_{crt}$ is the user's interest with the concept $crl$ in their profile.
- $crl$ is the $l^{th}$ concept among the most similar concepts with the result *r*.

The new score $new_{wt_r}$ is used to re-sort the documents.

Query enrichment, filtering results, and result reclassification are all techniques used in information retrieval systems to improve the relevance and effectiveness of search results, but they serve different purposes and operate at different stages of the retrieval process. Query enrichment focuses on improving the expressiveness of the user's query to retrieve more relevant documents. Filtering results aims to remove irrelevant or low-quality content from the retrieved documents. Result reclassification involves refining the ranking or ordering of search results based on additional criteria beyond their initial relevance scores. These techniques can be used individually or in combination to enhance the overall effectiveness of IR systems.

While query enrichment, filtering results, and result reclassification offer valuable strategies for improving personalized information access, they each have their own strengths and limitations. Balancing these factors and carefully considering the specific needs and preferences of users is essential for designing effective and user-centric information retrieval systems. Additionally, ongoing research and innovation are needed to address the limitations and further enhance the capabilities of these techniques. Table 1 presents the strengths and limitations of query enrichment, filtering results, and result reclassification.

**Table 1. Strengths and limitations of query enrichment, filtering results, and result reclassification**

| Approach | Strengths | limitations |
|---|---|---|
| **Query Enrichment** | • Expandable retrieval scope<br>• Addresses vocabulary mismatch<br>• User-centric approach | • Noise introduction<br>• Semantic ambiguity<br>• Over-reliance on external resources |
| **Filtering Results** | • Relevance enhancement<br>• Improved user experience<br>• Customization and control | • Information loss<br>• Bias and subjectivity<br>• Trade-off between precision and recall |
| **Result Reclassification** | • Refined ranking<br>• Enhanced personalization<br>• Dynamic adaptation | • Algorithm complexity<br>• User privacy concerns<br>• Feedback Loop bias |

# MOTIVATION AND METHODOLOGY

Personalized IR focuses on integrating the user profile into one of the phases of the information search process to better meet user needs. The modeling of user profiles is indeed at the heart of personalized IR. Personalization in IR aims to tailor search results and recommendations to individual users based on their preferences, behavior, and historical interactions with the system. In the context of personalized IR based on deep neural networks, a user profile can be defined as a representation of an individual user's preferences, interests, and behavior captured and modeled through the use of deep learning techniques. The user profile is essentially a learned embedding or set of parameters within a deep neural network that encodes the user's interactions with the system, historical preferences, and relevant contextual information.

Our contribution concerns the building of user profiles based on deep neural networks. We present different models based on these architectures for two complementary phases in order to infer a user profile. The profile definition will be built by its center of interest. We will use documents deemed implicitly relevant by a user (click + reading time) to automatically extract the domain of each document. This domain may correspond to a user's center of interest. The idea is to be able to make the best use of the content of the document in order to define its domain using deep networks. The latter have shown their effectiveness in numerous fields of application such as image processing, text categorization and medical diagnosis thanks to their capacity for classification and generalization. In our case, optimal deep network models are used to classify the text of a document according to a category. The process is divided into two phases: the first phase consists of deducing the domain of the document deemed implicitly relevant by a user which corresponds to their center of interest, for this we have created three models with different architectures. The second phase consists of using the domain predicted by the best model from the first phase with other characteristics of the user in order to predict other centers of interest.

The following subsections examine the two phases in the proposed approach: deducing document domains and predicting user centers of interest.

## DEDUCING DOCUMENT DOMAINS

The first phase allows us to deduce the domains of the documents. The predicted domain corresponds to the user's center of interest. For the modeling of this phase, we proposed three different architectures. Each of them is characterized by:

- The type of neural networks used such as ANNs (Artificial Neural Networks), CNNs (Convolutional Neural Networks), and RNN-LSTM (Recurrent Neural Networks- Long Short-Term Memory).
- The number of hidden layers.
- The number of units in each layer.

ANNs serve as the basis for more complex architectures like CNNs and RNNs. They are suitable for our approach since input features are relatively simple and independent. The use of ANN can model complex non-linear relationships between input and output variables. CNNs leverage shared weights and local connectivity to efficiently extract hierarchical features from input data. RNNs are specifically designed for sequential data processing tasks, where the order of input elements such as NLP, time-series analysis, and speech recognition matters. They maintain a state vector that captures information from previous time steps, allowing them to model temporal dependencies and capture long-range context effectively. RNN variant such as LSTM address the vanishing gradient problem and enable learning of long-range dependencies. RNNs are well-suited for tasks involving sequential or temporal data, where capturing dependencies between input elements is crucial. They can handle variable-length inputs and effectively model long-range dependencies, making them suitable for tasks like machine translation, sentiment analysis, and time-series prediction.

The selection of ANN, CNN, or RNN depends on the nature of the data, the task requirements, and the desired outcomes for each proposed model.

For the first model, we propose an ANN architecture as described in Figure 1.

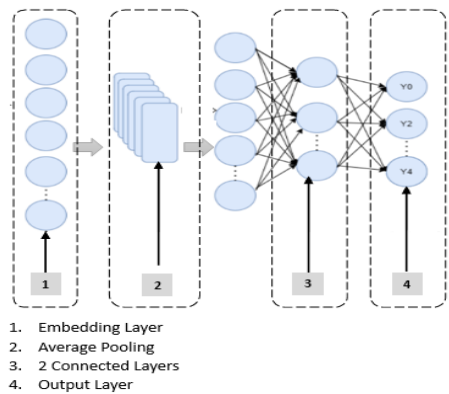

1. Embedding Layer
2. Average Pooling
3. 2 Connected Layers
4. Output Layer

**Figure 1. First model architecture**

This model is composed of an input layer, an average pooling layer, two fully connected layers and an output layer. Each of the previous layers has a *ReLU* or *SoftMax* activation function. The first layer of our network is an embedding layer which encodes the input text. It represents a description of size 200 in a sequence of dense vectors of dimension 64. The description size is set to 200. This value represents the size of the longest description in our dataset. The average pooling layer allows you to take the average of the input patch. The other two hidden layers have respectively 130 neurons and 70 neurons where the activation function used is *ReLU*. The last layer uses the *SoftMax* activation function which calculates the probability distribution of the five classes. The number of neurons is decreasing, this allows us to have a gradual transition from a high number of input neurons to 5 output neurons, to facilitate training and without making the network too complex.

The second model shown in Figure 2 is a simple architecture of an LSTM network.

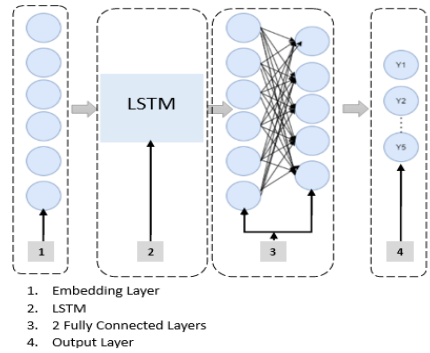

1. Embedding Layer
2. LSTM
3. 2 Fully Connected Layers
4. Output Layer

**Figure 2. Second model architecture**

The first layer of this network is an embedding layer that encodes the input text that represents a description of size 200 into a sequence of dense vectors of dimension 64. The next layer is a Long Short-Term Memory (bidirectional) layer which transforms our input into a single vector containing information about the entire sequence. It uses 128 neurons, the result obtained is then passed to the two fully connected layers of 128 and 64 neurons respectively. The output result is passed to the output layer using 5 neurons where the activation function is *SoftMax*.

The third model that we present in Figure 3 is a CNN architecture. It is composed of an embedding layer, a convolution layer, an average pooling layer and a fully connected layer, and the output layer. The first layer of our network is an embedding layer which encodes the input text, and which represents a description of size 200, in a sequence of dense vectors of dimension 64, the second layer is the convolution layer with 128 neurons, using a convolution window of size 5. The result is then passed to an average pooling layer in order to compress the information by reducing the size of our input. It uses 128 neurons and a pooling window of size 2, the output result will be transmitted to a fully connected layer of 64 neurons. The last layer uses the *SoftMax* activation function which calculates the probability distribution of the 5 classes.

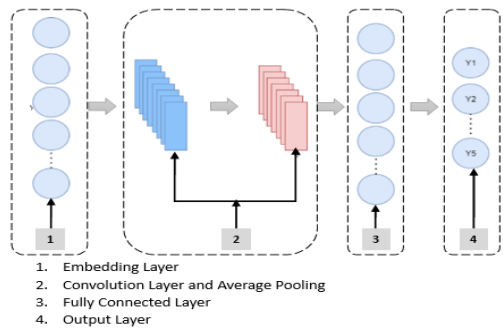

1. Embedding Layer
2. Convolution Layer and Average Pooling
3. Fully Connected Layer
4. Output Layer

**Figure 3. Third model architecture**

## PREDICTING USER CENTERS OF INTEREST

This second phase makes it possible to exploit the domain predicted by the best model of the first phase with other characteristics of the user such as age, gender, salary, and geography in order to predict other centers of interest. For example, if we have a more general center of interest such as sport, we will deduce a sub-center of interest, for example football, from existing profiles. For the first model we propose an ANN architecture described in Figure 4.

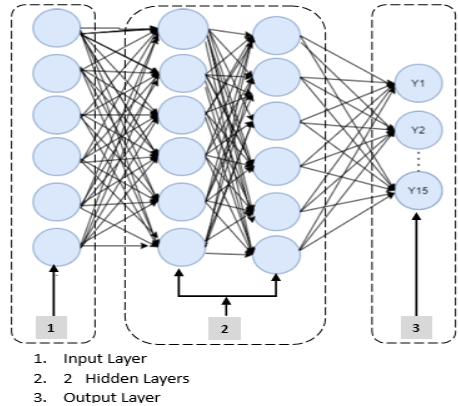

**Figure 4. Architecture of the first model for the second phase.**

The network consists of an input layer of six units, two hidden layers which have six units with a *ReLU* activation function. The output result is passed to the output layer using 15 neurons where the activation function is *SoftMax*. For the second model we propose an ANN architecture like the architecture described in figure 4. The network consists of an input layer of 50 neurons and two hidden layers which have 40 and 20 neurons respectively with an activation function *ReLU*. The output result is passed to the output layer using 15 neurons where the activation function is *SoftMax*.

# DATA PREPROCESSING AND REPRESENTATION

This section details data preprocessing and representation models used in the learning-based models.

## DATASETS

For the validation of our approach, we used two collections of data. The first collection brings together a public BBC dataset made up of 2225 articles, organized into five folders: *Business*, *Entertainment*, *Politics*, *Sport*, and *Technology*. The dataset presents a valuable resource for researchers and practitioners interested in NLP and information retrieval, offering a diverse collection of articles across different categories for analysis and experimentation (Bose, 2019).

To use this collection, we have labeled each article by its corresponding category. The result is presented in Table 2.

**Table 2. Structure of the first collection**

| Article # | Category | Text |
|---|---|---|
| 1 | TV future in the hands of viewers with home theatre systems plasma high-definitions TVs and digital videos recorders moving into the living room the way people watch TV will be radically different in five years' time.... | Technology |
| 2 | Cars pull down US retail figures ... | Business |
| 3 | Joy Division story to become film ... | Entertainment |
| 4 | Vera Drake scoops film award ... | Entertainment |
| 5 | Media seek Jackson 'juror' notes ... | Entertainment |
| 6 | Budget to set scene for election ... | Politics |
| 7 | Howard denies split over ID cards ... | Politics |
| 8 | Kerr frustrated at victory margin.... | Sport |
| 9 | Chepkemei joins Edinburgh line-up .... | Sport |
| 10 | Wi-fi web reaches farmers in Peru ... | Technology |
| 11 | Seamen sail into biometric future ... | Technology |
| … | … | … |
| 2225 | Seamen sail into biometric future ... | Technology |

We have divided the data in our collection into two parts:

- Training data: which constitutes 80% of our dataset, namely 1780 documents. This data is used to generate the learning model.
- Test data: which constitutes the remaining 20% of our dataset, namely 445 documents. The learning model will then be applied to this data for testing purposes.

The second collection brings together a set of 10,000 records. They are represented in the form of columns: *User-ID*, *User*, *Location*, *Gender*, *Age*, *Domain*, *Salary*, and *Center of Interest*. Table 3 presents the sample data from this collection except *User-ID* and *User* features. We have divided the data in our collection into two parts:

- Training data: which constitutes 80% of our dataset, namely 8000 profiles. This data is used to generate the learning model.
- Test data: which constitutes the remaining 20% of our dataset, namely 2000 profiles. The learning model will then be applied to this data for testing purposes.

**Table 3. Structure of the second collection**

| # | Location | Gender | Age | Domain | Salary | Center of Interest |
|---|----------|--------|-----|--------|--------|--------------------|
| 1 | France | Female | 41 | Technology | 101348.88 | Computer Scientist |
| 2 | Spain | Female | 42 | Sport | 113931.58 | Arbitrator |
| … | … | … | … | … | … | … |
| 12 | Spain | Male | 24 | Entertainment | | Actor |
| … | … | … | … | … | … | … |

## PREPROCESSING AND REPRESENTATION

For the preprocessing of the data from the first collection, we cleaned the documents and their contents of all unnecessary words and characters. Then we applied tokenization. After completing the preprocessing task, we obtained labeled text documents. To maintain syntactic and semantic similarity, word embeddings project vocabulary terms into a low-dimensional space. Consequently, words must be semantically or syntactically near if their distance vectors are close to one another. Word latent features, which can capture syntactic and semantic aspects, are represented by each dimension.

Several neural approaches have been proposed in the literature for the construction of word embeddings, including GloVe (GLObal VEctor) (Pennington et al., 2014). GloVe is a dictionary that associates a vector with each word. To do this, it collects the co-occurrence characteristics of words in the form of a matrix, to construct compacted representations of the documents. It combines *Count-based* matrix factorization with predictive or neural models, as shown by (Levy et al., 2015; J. Li & Jurafsky, 2015; Pennington et al., 2014) that views this method as a predictive model, whereas (Arora et al., 2016) views it as a *Count-based* model. Its foundation is the creation of a global co-occurrence matrix, called *GM*, of words using a sliding contextual window to process the corpus.

The number of times the word $m_j$ appears in relation to the term $m_i$ is represented by each element $GMij$ in this instance. GloVe is an unsupervised learning model that considers all of the data that the corpus contains, not only the data that is included inside a word window. A least squares regression model is trained to create the vector representations $\vec{m_i}$ and $\vec{m_j}$ once the matrix *GM* has been computed. The following important details on the co-occurrence of word pairs $m_i$ and $m_j$ must be retained in these representations:

$$\overrightarrow{m_i}^T \overrightarrow{m_j} \;+\; b_i \;+\; b_j = \log\big(GM_{ij}\big) \tag{2}$$

where the corresponding bias vectors for the words $m_i$ and $m_j$ are $b_i$ and $b_j$, respectively. The process involves reducing the cost function, which assesses the total squared errors:

$$E = \sum_{i,j=1}^{nv} f\big(GM_{ij}\big)(\overrightarrow{m_i}^T \overrightarrow{m_j} \;+\; b_i \;+\; b_j - \log\big(GM_{ij}\big))^2 \tag{3}$$

where $f(.)$ is a weighting function that weights the cost based on the frequency of the co-occurrence number $GMij$, where nv is the vocabulary size. This has the following definition:

$$f\big(GM_{ij}\big) = \begin{cases} (\dfrac{GM_{ij}}{GM_{max}})^{\alpha} \; if \; GM_{ij} < GM_{max} \\ 1 \qquad\qquad\qquad else \end{cases} \tag{4}$$

where $\alpha = 3/4$ and $GM_{max} = 100$. The function returns 1 in the case that the co-occurrence value $GM_{ij}$ of a pair of words is exceptionally high, meaning that it exceeds the maximum value $GM_{max}$. If not, it transmits a weight between 0 and 1 to the other pairings, with alpha determining how the weights are distributed in this range. These processing steps are depicted in Figure 5.

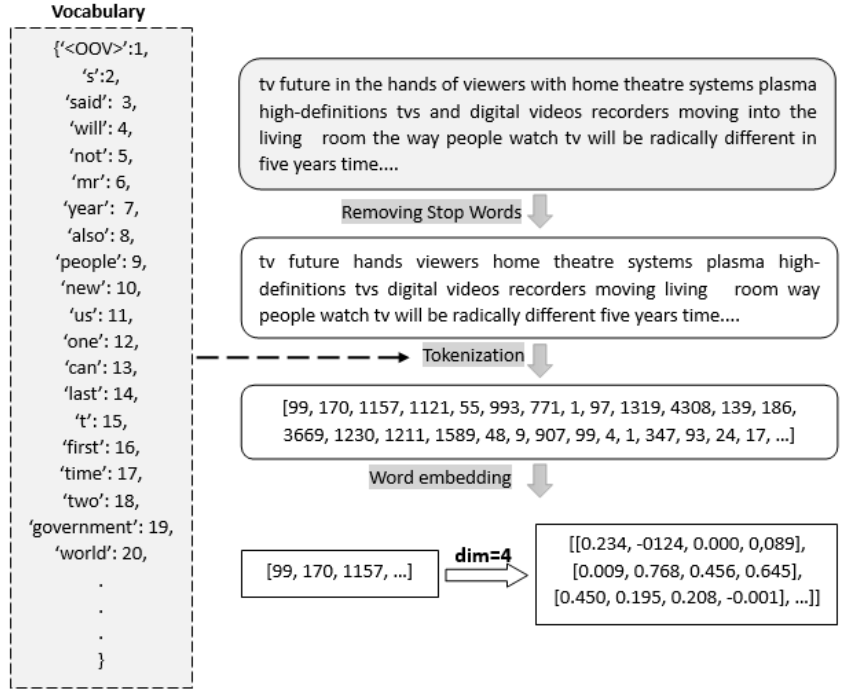

**Figure 5. Preprocessing and representation of the first collection**

For the preprocessing and representation of data from the second collection (Location, Gender, Age, Domain, Salary, and Center of Interest), we transformed them into digital form. These values will be represented as a vector which will be ready for the training phase. Figure 6 shows the processing processes in detail.

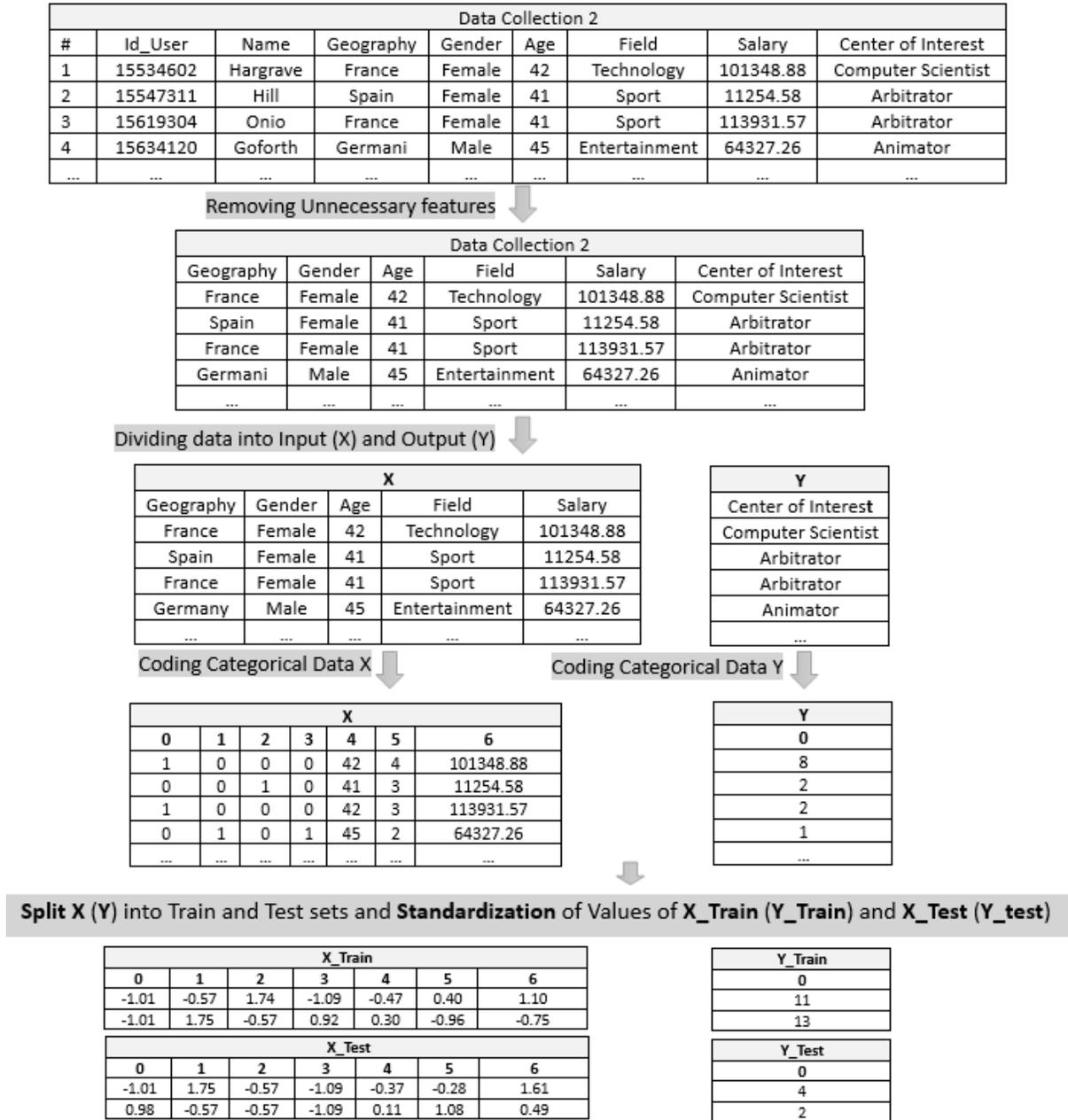

**Figure 6. Preprocessing and representation of the second collection**

# COMPUTATIONAL RESULTS

The purpose of this section is to explore and evaluate the results of our models during the two phases: deducing document domains and Predicting user centers of interest.

## MODEL EVALUATION

When a model is developed, it is important to be able to evaluate its functioning and its ability to meet the objectives set for it. In deep learning, loss and accuracy are two crucial metrics used for evaluating the performance of a model during training and testing phases.

- Accuracy: Accuracy is the most used measure to judge and evaluate the performance of a model. It is defined by the following formula (Jierula et al., 2021; Sassi Hidri et al., 2022):

$$Accuracy = \frac{Number\ of\ CorrectPredictions}{Total\ Number\ of\ Predictions\ made} \tag{5}$$

- Loss: Loss, also known as the cost function or objective function, quantifies the difference between the predicted output and the actual target labels indicates how well a given model performs after each optimization iteration. It provides a measure of how well the model is performing on the training data. The goal is to minimize this loss during the training process Ideally, there should be a reduction in losses after each or more iterations (Wang et al., 2022). It is then necessary to know how to choose the loss function, otherwise the model risks never doing what we ask it to do. Since we are facing a multi-class classification problem and the activation function chosen in the last layer of our models is *SoftMax*, we have chosen the categorical cross-entropy function to measure the loss of the models. It is defined by the following formula (P. Li et al., 2021):

$$Loss = -\sum_{j=1}^{K} y_j \log(\hat{y}_j) \tag{6}$$

where $K$ is classes, $y$ is the accrual value, and $\hat{y}$ in the neural network prediction.

## DEDUCING DOCUMENT DOMAINS

We trained our three models described in the previous section on the training data set from the first collection. Subsequently, we tested the three models on the test dataset. We show in the following the results obtained by carrying out an evaluation in terms of accuracy and loss. The first model (RNN) is built with the configuration shown in Table 4.

**Table 4. $1^{st}$ model architecture**

| Lyer (Type) | Output Shape | Param |
|---|---|---|
| embedding_1 (Embedding) | (None, None, 64) | 320000 |
| global_average_pooling1d_1 | (None, 64) | 0 |
| dense_2 (Dense) | (None, 130) | 8450 |
| dense_3 (Dense) | (None, 70) | 9170 |
| dense_4 (Dense) | (None, 6) | 426 |
| Total params: 338046 | | |
| Trainable params: 338046 | | |
| Non-trainable params: 0 | | |

`The training set's accuracy curve is displayed in blue in Figure 7, whereas the test set's accuracy is displayed in orange as a function of the number of epochs. The training set's loss curve is displayed in blue in Figure 8, whereas the test set's loss is displayed in orange as a function of the number of epochs.

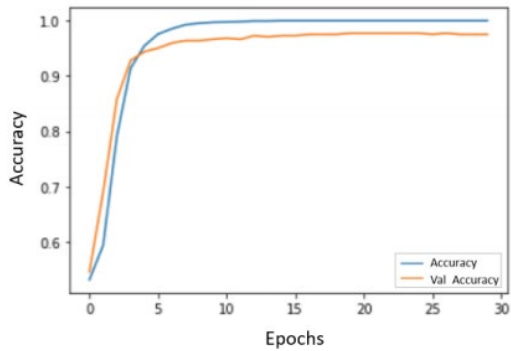 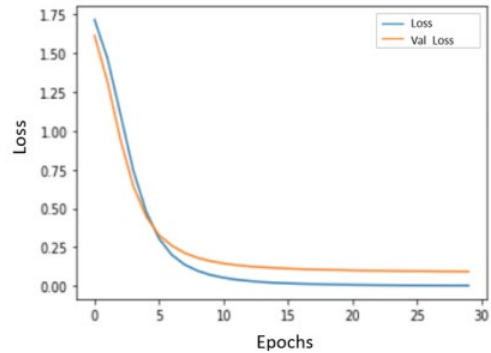

**Figure 7. 1$^{st}$ model accuracy**          **Figure 8. 1$^{st}$ model loss**

As shown in Figures 7 and 8, the number of epochs rises, so does the accuracy of training and validation. This illustrates how the model gains new knowledge with every era. The training and validation errors also get less as the number of epochs increases. The first model's findings are presented in Table 5.

**Table 5. Results obtained for the 1$^{st}$ model**

| Accuracy with training set | Accuracy with test set |
|---|---|
| 99.49% | 97.08% |

The second model is built with the configuration shown in Table 6.

**Table 6. 2$^{nd}$ model architecture**

| Lyer (Type) | Output Shape | Param | |
|---|---|---|---|
| embedding (Embedding) | (None, None, 64) | 320000 | |
| bidirectional (Bidirectional) | (None, 128) | 66048 | |
| dense (Dense) | (None, 128) | 16512 | |
| dense_1 (Dense) | (None, 64) | 8256 | |
| dense_2 (Dense) | (None, 6) | 390 | |
| Total params: 411206 | | | |
| Trainable params: 411206 | | | |
| Non-trainable params: 0 | | | |

The training set's accuracy curve is shown in blue in Figure 9, whereas the test set's accuracy is shown in orange as a function of the number of epochs. The training set's loss curve is shown in blue in Figure 10, whereas the test set's loss is shown in orange as a function of the number of epochs.

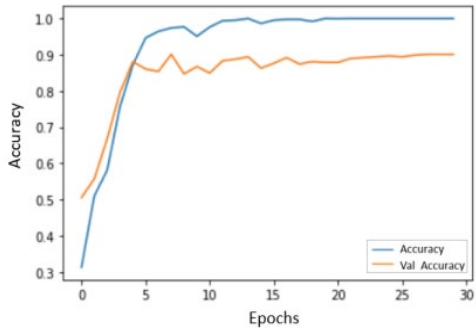
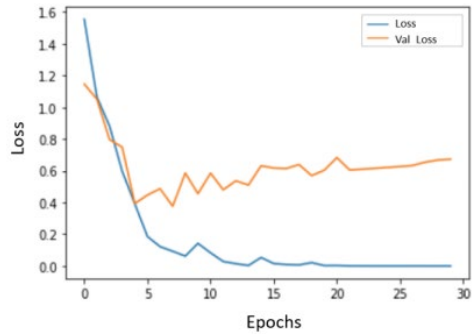

**Figure 9.** 2$^{nd}$ **model accuracy**          **Figure 10.** 2$^{nd}$ **model loss**

Figures 9 and 10 show that as the number of epochs rises, so does the accuracy of training and validation. This illustrates how the model gains new knowledge with every era. The training and validation errors also get less as the number of epochs increases. However, we find that there is overfitting, with a 10% discrepancy between the test pressure and training accuracy. The findings are displayed in Table 7.

**Table 7. Results obtained for the 2$^{nd}$ model**

| Accuracy with training set | Accuracy with test set |
|---|---|
| 99.20% | 90.01% |

After the results obtained with the second model and in order to improve the learning rate and avoid overfitting, we added a regularizer (Dropout). Its configuration is shown in Table 8.

**Table 8. 2$^{nd}$ model variant architecture**

| Lyer (Type) | Output Shape | Param # |
|---|---|---|
| embedding_2(Embedding) | (None, None, 64) | 320000 |
| bidirectional_2 (Bidirectional) | (None, 128) | 66 048 |
| dense_6 (Dense) | (None, 128) | 16512 |
| dropout_2 (Dropout) | (None, 128) | 0 |
| dense_7 (Dense) | (None, 64) | 8256 |
| Total params: 369734 | | |
| Trainable params: 369734 | | |

The training set's accuracy curve is displayed in blue in Figure 11, whereas the test set's accuracy is displayed in orange as a function of epoch number. The training set's loss curve is shown in blue in Figure 12, whereas the test set's loss is shown in orange as a function of epoch number.

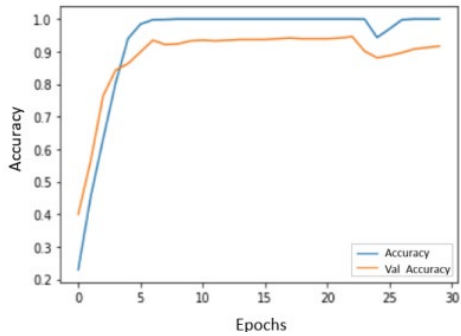 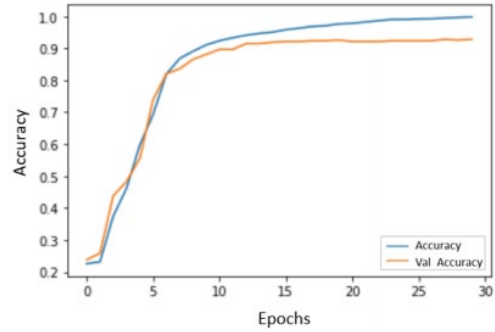

**Figure 11. 2$^{nd}$ model variant accuracy**          **Figure 12. 2$^{nd}$ model variant loss**

As shown in Figures 11 and 12, learning and validation accuracy is higher than that of the second model. We discover that the accuracy dropped after reaching 94% at epoch 23. Thus, we conclude that the model's performance is influenced by the number of epochs. The findings obtained with this model are displayed in Table 9.

**Table 9. Results obtained for the variant of the 2$^{nd}$ model**

| Accuracy with training set | Accuracy with test set |
|---|---|
| 99.50% | 92.07% |

The third model is built with the configuration shown in table 10.

**Table 10. 3$^{rd}$ model architecture**

| Lyer (Type) | Output Shape | Param # |
|---|---|---|
| embedding_1(Embedding) | (None, None, 64) | 320000 |
| conv1d_1 (Conv1d) | (None, None, 128) | 41088 |
| global_average_pooling1d (Global) | (None, 128) | 0 |
| dense_2 (Dense) | (None, 64) | 8256 |
| dense_3 (Dense) | (None, 6) | 390 |
| Total params: 369734 | | |
| Trainable params: 369734 | | |
| Non-trainable params: 0 | | |

The training set's accuracy curve is shown in blue in Figure 13, whereas the test set's accuracy is shown in orange as a function of epoch number. The training set's loss curve is shown in blue in Figure 14, whereas the test set's loss is shown in orange as a function of epoch number.

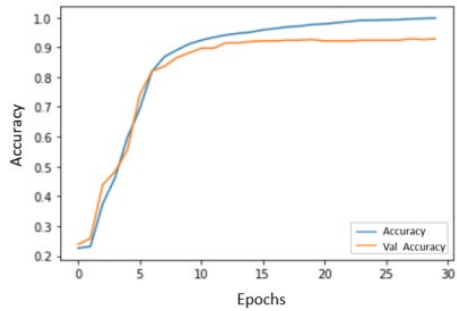

**Figure 13.** $3^{rd}$ **model accuracy**

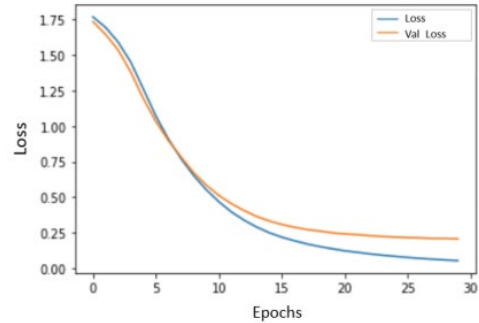

**Figure 14.** $3^{rd}$ **model loss**

As shown in Figures 13 and 14, as the number of epochs rises, so does the accuracy of training and validation. This illustrates how the model gains new knowledge with every era. The training and validation errors also get less as the number of epochs increases.

Table 11 shows the results obtained for the third model.

**Table 11. Results obtained for the variant of the $3^{rd}$ model**

| Accuracy with training set | Accuracy with test set |
|---|---|
| 99.80% | 97.50% |

The results obtained for the three models are compared in Table 12.

**Table 12. Comparison of the results obtained from the three models with the first collection**

| Model # | Epochs | Accuracy with training set | Accuracy with test set |
|---|---|---|---|
| Model 1 | 30 | 99.49% | 97.08% |
| Model 2 | 30 | 99.20% | 90.01% |
| Variant Model 2 | 30 | 99.50% | 92.07% |
| Model 3 | 30 | 99.80% | 97.50% |

We observe that the results obtained are quite similar for the three models (1, 2 and 3). We see that different parameters influence the performance of the model, namely the architecture of the networks, the number of epochs, the regularization which makes it possible to avoid overfitting. According to these results, we see that the third model is the most efficient.

## PREDICTING USER CENTERS OF INTEREST

The first model is built with the configuration shown in Table 13.

**Table 13. $1^{st}$ model architecture**

| Lyer (Type) | Output Shape | Param |
|---|---|---|
| dense_1 (Dense) | (None, 6) | 42 |
| dense_2 (Dense) | (None, 6) | 42 |
| dense_3 (Dense) | (None, 6) | 42 |
| dense_4 (Dense) | (None, 15) | 105 |
| Total params: 231 | | |
| Trainable params: 231 | | |
| Non-trainable params: 0 | | |

The training set's accuracy curve is displayed in blue in Figure 15, whereas the test set's accuracy is displayed in orange as a function of epoch number. The training set's loss curve is shown in blue in Figure 16, whereas the test set's loss is shown in orange as a function of epoch number.

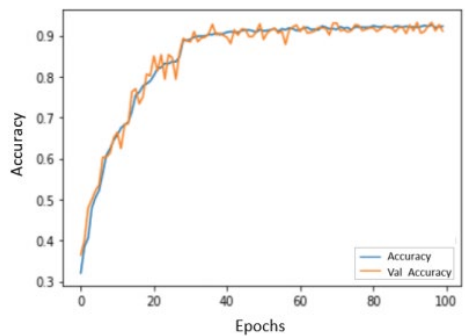
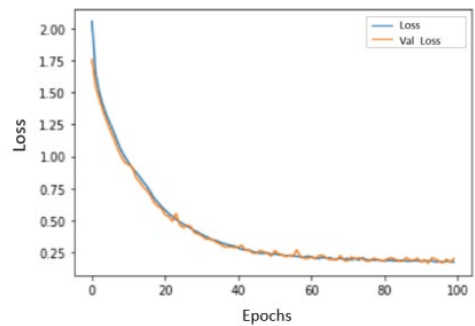

**Figure 15. $1^{st}$ model accuracy**        **Figure 16. $1^{st}$ model loss**

As shown in Figures 15 and 16, as the number of epochs rises, so does the accuracy of training and validation. This illustrates how the model gains new knowledge with every era. The training and validation errors also get less as the number of epochs increases. However, the learning outcome is unsatisfactory.

Table 14 shows the results obtained for the first model.

**Table 14. Results obtained for the $1^{st}$ model**

| Accuracy with training set | Accuracy with test set |
|---|---|
| 79.85% | 78.75% |

The second model is built with the configuration shown in table 15.

**Table 15. $2^{nd}$ model architecture**

| Lyer (Type) | Output Shape | Param |
|---|---|---|
| dense_12 (Dense) | (None, 50) | 300 |
| dense_13 (Dense) | (None, 40) | 2040 |
| dense_14 (Dense) | (None, 20) | 820 |
| dense_15 (Dense) | (None, 15) | 315 |
| Total params: 3,475 | | |
| Trainable params: 3,475 | | |
| Non-trainable params: 0 | | |

The training set's accuracy curve is shown in blue in Figure 17, whereas the test set's accuracy is shown in orange as a function of epoch number. The training set's loss curve is shown in blue in Figure 18, whereas the test set's loss is shown in orange as a function of epoch number.

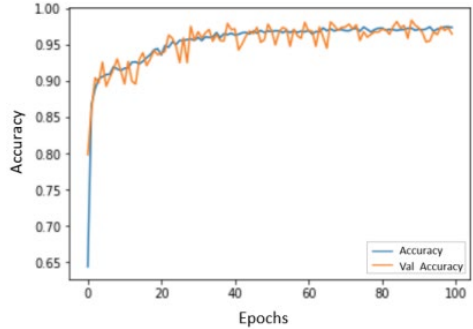 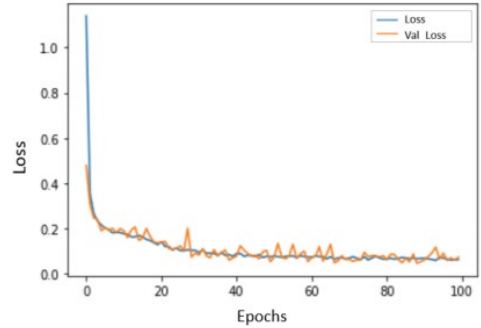

**Figure 17. $2^{nd}$ loss accuracy**          **Figure 18. $2^{nd}$ loss**

As shown in Figures 17 and 18, as the number of epochs rises, so does the accuracy of training and validation. This illustrates how the model gains new knowledge with every era. The training and validation errors also get less as the number of epochs increases. From the results obtained, we can deduce that the larger the network, the better the algorithm learns and predicts.

Table 16 shows the results obtained for the third model.

**Table 16. Results obtained for the $2^{nd}$ model**

| Accuracy with training set | Accuracy with test set |
|---|---|
| 97.39% | 96.50% |

The results obtained for the two models are compared in Table 17.

**Table 17. Comparison of the results obtained from the two models with the second collection**

| Model # | Epochs | Accuracy with training set | Accuracy with test set |
|---|---|---|---|
| Model 1 | 100 | 79.85% | 78.75% |
| Model 2 | 100 | 87.39% | 96.50% |

From these results, we can see that using a larger architecture allows the model to learn better and predict better. The experiments showed that the third model of the first collection and the second model of the second collection are ideal to have a better training and testing result.

# CONCLUSIONS

The objectives of learning-based models for building user profiles for personalized information access revolve around delivering personalized, relevant, and engaging experiences to users while also driving positive outcomes for businesses and organizations. These models aim to improve the overall user experience by providing personalized recommendations and tailored content that match the interests, preferences, and needs of individual users. By building accurate user profiles, they can deliver relevant and engaging information to users, leading to increased satisfaction and engagement.

In this paper, the user profile is defined by its center of interest which we automatically infer using deep networks. To automatically infer interests, we proposed different deep network models that we evaluated using data collections. Firstly, we inferred a center of interest from documents from the BBC collection. Once the center of interest is inferred, we enrich it with other more specific centers of interest, using it as input to a neural network with other user-related data from the collection we have created.

The different experiments carried out showed the interest of the models adopted, with a very highest result in the third model of the first collection and the second model of the second collection.

Despite the effectiveness of learning-based models, their black-box nature can hinder understanding and trust. Future research could focus on developing methods to enhance the interpretability of these models, enabling users to understand how their profiles are constructed and personalized recommendations are generated. Developing robust evaluation methodologies for assessing the effectiveness and fairness of learning-based models for building user profiles is essential for advancing research in this area. Future directions might involve developing standardized benchmarks, metrics, and evaluation protocols for comparing different profiling approaches and measuring their impact on user satisfaction and information access outcomes. Moreover, increasing the interpretability and transparency of learning-based models for building user profiles is essential for building user trust and understanding model decisions. Future directions could also involve developing interpretable machine learning techniques and visualization tools to explain how user profiles are constructed and used for personalized information access.

# AUTHOR

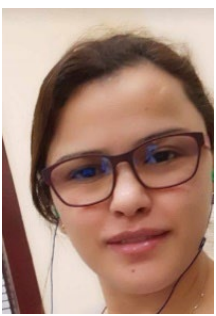

**Minyar Sassi Hidri** was born in Nabeul, Tunisia. She received an Engineering Degree in Computer Sciences and Ph.D. from the National Engineering School of Tunis (ENIT), University of Tunis El Manar (UTM), Tunisia, in 2003 and 2007, respectively. She obtained the qualification to lead research in computer sciences from the UTM, Tunisia, in June 2014. She is currently an Assistant Professor at the Imam Abdulrahman Bin Faisal University, Dammam, Saudi Arabia. She has been teaching computer science and information systems for more than 20 years. Her research interests include combinatorial aspects in Big Data analytics, machine learning, deep learning, and text mining, with over 70 publications. She is also a member of the steering committee of many international conferences and a reviewer of impacted journals. You can contact her via email at mmsassi@iau.edu.sa